\documentclass{ifacconf}

\usepackage{graphicx}      
\usepackage{natbib} 
\usepackage{amsmath}
\usepackage{amssymb}     

\usepackage{subfigure}   

\newtheorem{remark}{Remark}  
       
\begin{document}
\begin{frontmatter}

\title{Seasonalities and cycles in time series: \\ A fresh look with computer experiments 
} 


\author[First,Third]{Michel Fliess} 
\author[Second,Third,Fourth]{C\'{e}dric Join}

\address[First]{LIX (CNRS, UMR 7161), \'Ecole polytechnique, 91128 Palaiseau, France (e-mail: Michel.Fliess@polytechnique.edu)}
\address[Second]{CRAN (CNRS, UMR 7039), Universit\'{e} de Lorraine, BP 239, 54506 Vand{\oe}uvre-l\`{e}s-Nancy, France \\ (e-mail: cedric.join@univ-lorraine.fr)}
\address[Third]{AL.I.E.N. (ALg\`{e}bre pour Identification \& Estimation Num\'{e}riques), 24-30 rue Lionnois, BP 60120, 54003 Nancy, France \\ (e-mail: \{michel.fliess, cedric.join\}@alien-sas.com)}
\address[Fourth]{Projet Non-A, INRIA Lille -- Nord-Europe, France}

\begin{abstract}                
Recent advances in the understanding of time series permit to clarify seasonalities and cycles, which might be rather obscure in today's literature. A theorem 
due to P. Cartier and Y. Perrin, which was published only recently, in 1995, and several time scales yield, perhaps for the first time, a clear-cut definition of seasonalities and cycles.  
Their detection and their extraction, moreover, become easy to implement. Several 
computer experiments with concrete data from various fields are presented and discussed. The conclusion suggests the application of this approach to the debatable Kondriatev waves.
\end{abstract}


\end{frontmatter}

\section{Introduction}\label{intro}
Everyone who is interested in econometrics, finance, sales, various questions related to management, and in many other topics, like, for instance, the study of health and climate data (see, \textit{e.g.}, \cite{health,mudel}), has heard the words \emph{seasonalities} and \emph{cycles}, as well as the word \emph{deseasonalization}.  Although seasonalities, cycles, their detection and their extraction may be found since ever in many publications (see, \textit{e.g.}, \cite{baum,bourb,brock,geman,ghy,holt,levin,melard,miron,tsay}, and the references therein), this subject, which plays such an fundamental r\^{o}le in not only concrete but also abstract questioning, seems to be far from being fully understood (see, \textit{e.g.}, \cite{us}). A lack of any clear-cut definition might be the explanation. Seasonalities and cycles are seldom strictly periodic phenomena (see, \textit{e.g.}, \cite{bourb,fran}), which could be then detected and extracted thanks to classic means like Fourier techniques stemming from signal processing.  The actual situation which might be summarized from today's literature is rather obscure: 
\begin{itemize}
\item Seasonalities are often, but not always, related to time horizons which are shorter than one year. Time horizons for cycles are often, but not always, longer.
\item Seasonalities are often, but not always, assumed to exhibit a ``more'' periodic behavior than cycles.
\end{itemize}
Tedious \textit{ad hoc} approaches are therefore developed for concrete case-studies (see, \textit{e.g.}, \cite{boro}, \cite{hamilton0}, \cite{health}, \cite{satel}, \cite{koop}, \cite{zhang}, several contributions in \cite{us}, and the references therein).

The following additive decomposition is often encountered (see, \textit{e.g.}, \cite{bourb,brock,ghy,melard}):
\begin{equation}\label{decomp}
\mu_t + S_t + C_t + R_t
\end{equation}
where
\begin{enumerate}
\item $\mu_t$ is the \emph{trend},
\item $S_t$ is the seasonal component,
\item $C_t$ is the cyclic component,
\item $R_t$ is the \emph{residual} component.
\end{enumerate}
Cycles are sometimes ignored. It yields to
\begin{equation}\label{decompo}
\mu_t + S_t + R_t
\end{equation}
From a mathematical standpoint and to the best of our knowledge,  the decompositions \eqref{decomp} and \eqref{decompo} have never been rigorously proved. 
\begin{remark}
This is confirmed by the fact that 
\begin{itemize}
\item mathematically oriented treatises on time series, like for instance those by \cite{hamilton} and by \cite{gourieroux}, 
\item the historical and epistemological study due to \cite{meuriot}
\end{itemize}
do not mention seasonalities and cycles in spite of their indubitable practical usefulness.
\end{remark}
Our aim in this communication is to suggest another route and to illustrate it via some computer experiments. Two main ingredients are employed:
\begin{enumerate}
\item Tools stemming from Robinson's \emph{nonstandard analysis} (see, \textit{e.g.}, \cite{diener,bams,robinson}) are used for a new theory of time series (see \cite{perp,esaim,agadir}, and the references therein).\footnote{Besides financial engineering 
this approach has also been utilized for short-term meteorological and road traffic forecasts (see \cite{paris} and \cite{troyes}).} A rather recent theorem due to \cite{cartier} yields a new understanding of their structure. 
\item Several time scales.
\end{enumerate}
Detection and extraction of seasonalities and cycles become much simpler. The quite involved techniques, like nonstationary time series modeling, which are often employed in the literature, become pointless in our \emph{model-free} setting (see also \cite{ijc} for further 
comments on this viewpoint).

Our communication is organized as follows. Section \ref{TS} is sketching our approach to time series. Seasonalities, cycles and deseasonalization are defined in Section \ref{season}.
Section \ref{computer} shows and discusses computer experiments. Let us emphasize the last example, in Section \ref{internet}, where the introduction of several time scales, which was suggested in Section \ref{time scales}, 
turns out to be most enlightening. Some concluding comments, with a hint on the celebrated but controversial Kondriatev waves, may be found in Section \ref{conclusion}.

\section{Revisiting time series}\label{TS}
\subsection{Time series and nonstandard analysis}
Take a time interval $[0, 1]$. Introduce as often in
nonstandard analysis the infinitesimal sampling
\begin{equation}\label{sampling}
T = \{ 0 = t_0 < t_1 < \dots < t_\nu = 1 \}
\end{equation}
where $t_{i+1} - t_{i}$, $0 \leq i < \nu$, is {\em infinitesimal},
{\it i.e.}, ``very small.'' A time series\footnote{See (\cite{perp,esaim,agadir}) for more details.} $X$ is a function
$T \rightarrow \mathbb{R}$.
\begin{remark} \label{rem}
Infinitely small or large numbers should be understood as mathematical idealizations. In practice a time lapse of $1$ second (resp. hour) should be viewed as quite small when compared to $1$ hour (resp. month). Nonstandard analysis may therefore be applied in concrete situations.
\end{remark}
\subsection{The Cartier-Perrin theorem}
A time series ${\mathcal{X}}: T \rightarrow \mathbb{R}$
is said to be {\em quickly fluctuating}, or {\em oscillating} around $0$, if,
and only if, its \emph{Lebesgue-integral} $\int_A {\mathcal{X}} dm$ (\cite{cartier}) is
infinitesimal for any \emph{appreciable} interval $A$, \textit{i.e.}, an interval which is neither infinitely small nor infinitely large.

According to a theorem due to \cite{cartier} the following additive decomposition holds for any time series $X$, which satisfies a weak integrability condition,
\begin{equation}\label{decomposition}
\boxed{X(t) = E(X)(t) + X_{\tiny{\rm fluctuation}}(t)}
\end{equation}
where 
\begin{itemize}
\item the \emph{mean}\footnote{ Also called \emph{average} or \emph{trend}. Let us emphasize that the notion of trend in the classic literature on time series is very different (see, \textit{e.g.}, \cite{bourb,brock,melard,tsay}).} 
$E(X)(t)$ is ``quite smooth,'' 
\item $X_{\tiny{\rm fluctuation}}(t)$ is quickly fluctuating.
\end{itemize}
Decomposition \eqref{decomposition}, which is unique up to an additive 
infinitesimal number, is replacing here Equations \eqref{decomp} and \eqref{decompo}.
\begin{remark}
The very definition of quick fluctuation shows that the mean $E(X)$ in Equation \eqref{decomposition} may be estimated via \emph{moving average} techniques, which are already popular in the time series literature.
\end{remark}

\section{Seasonalities and cycles}\label{season}
\subsection{Several time scales}\label{time scales}
From Remark \ref{rem}, taking different time scales becomes obvious. Assume for instance that the time series $X(t)$ is defined via a time scale where  $t_{i+1} - t_{i} = 1\  \text{second}$ in the sampling \eqref{sampling}. 
The two time series 
\begin{equation}\label{sumcumulee}
{\tt X}(t)  = \sum_{\iota = 1}^{\iota = 3600} X(t - \iota)
\end{equation}
and
\begin{equation*}
\textsf{ X}(t)  = \frac{\sum_{\iota = 1}^{\iota = 3600} X(t - \iota)}{3600}
\end{equation*}
are then defined with respect to a $1$ hour sampling.

\subsection{Some definitions}
Seasonalities and cycles are defined with respect to a rather ``coarse'' sampling \eqref{sampling}. Let us emphasize that different time scales might lead to different seasonalities and cycles (see Section \ref{internet}).. 

Rewrite Equation \eqref{decomposition} 
$$
X_{\tiny{\rm fluctuation}}(t)=X(t) -E(X)(t) 
$$
Introduce a threshold $\varpi > 0$ around the mean $E(X)$. A \emph{crossing time} $\mathcal{T}$ is defined by the two following properties:
\begin{enumerate}
\item $\mid X_{\tiny{\rm fluctuation}}({\mathcal T}) \mid  > \varpi$.
\item $\mid X_{\tiny{\rm fluctuation}}({\mathcal T} - h) \mid \leq \varpi$, where $h > 0$ is a sampling period.
\end{enumerate}
Let $\{ {\mathcal T}_1, \dots, {\mathcal T}_N \}$ be the set of crossing times where, for instance,  $X_{\tiny{\rm fluctuation}}({\mathcal T}) >  \varpi$. If ${\mathcal T}_{i + 1} - {\mathcal T}_i$ is ``approximately''periodic,\footnote{Introducing a threshold is an obvious necessity for defining this notion of approximate periodicity.} then we have a \emph{strong seasonality}, or a \emph{strong cycle}.\footnote{We speak of seasonality if ${\mathcal T}_{i + 1} - {\mathcal T}_i$ is less 
than 1 year, of cycle if not.} When there is no approximate periodicity,  we have a \emph{weak seasonality}, or a \emph{weak cycle}.
\subsection{Deseasonalization}
A deseasonalization of a time series $X$ is nothing else than replacing it by its mean $E(X)$ according to Equation \eqref{decomposition}. In other words fluctuations are removed.

\section{Computer illustrations}\label{computer}

Data for the first three examples are provided by the French \textit{Insee} (\textit{\underline{I}nstitut \underline{n}ational de la \underline{s}tatistique et des \underline{\'{e}}tudes \underline{\'{e}}conomiques}).\footnote{\tt http://www.insee.fr/fr/bases-de-donnees/} 
Consider three types of monthly data: 
\begin{itemize}
\item consumer prices of several foodstuffs in France,
\item job seekers registered at the French employment agency (\textit{P\^ole emploi}),
\item international passengers in Paris airports.
\end{itemize}
The internet traffic data of the last example, in Section \ref{internet},  are borrowed from \cite{Cortez}.

\subsection{Foodstuffs prices}

\begin{figure*}
\begin{center}
\subfigure[Consumer prices in France (--blue) and its mean (--red)]{
\resizebox*{9.1cm}{!}{\includegraphics{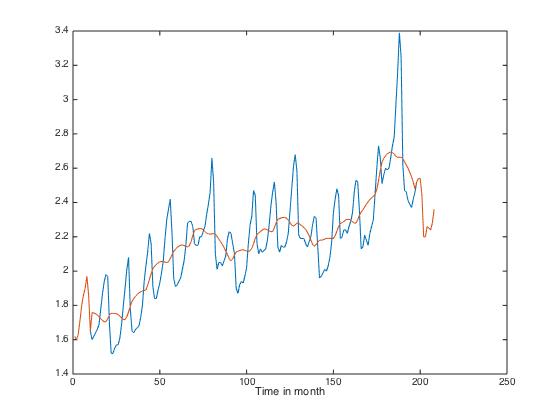}}}%
\subfigure[Fluctuations (--blue) and threshold (--red)]{
\resizebox*{9.1cm}{!}{\includegraphics{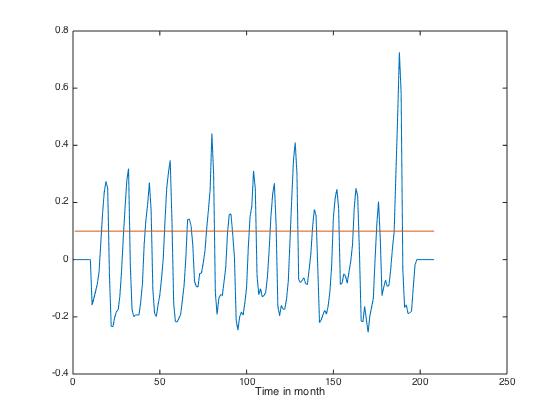}}}\\%
\subfigure[Time lapse between fluctuations]{
\resizebox*{9.1cm}{!}{\includegraphics{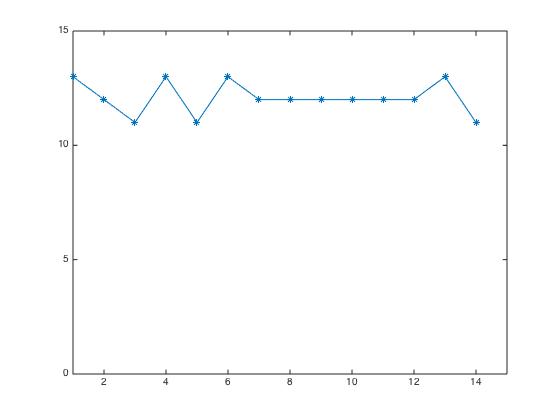}}}%
\caption{Apples (1 kg)}%
\label{P}
\end{center}
\end{figure*}

\begin{figure*}
\begin{center}
\subfigure[Consumer prices in France (--blue) and its mean (--red)]{
\resizebox*{9.1cm}{!}{\includegraphics{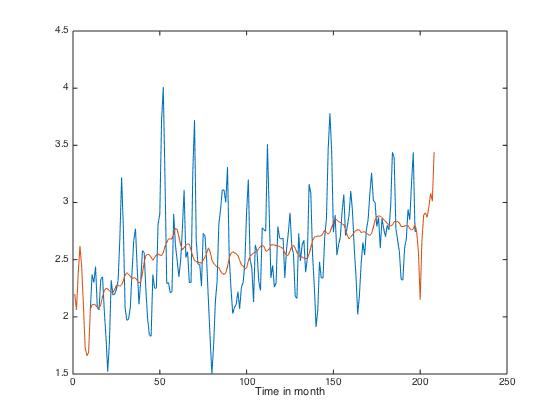}}}%
\subfigure[Fluctuations (--blue) and threshold (--red)]{
\resizebox*{9.1cm}{!}{\includegraphics{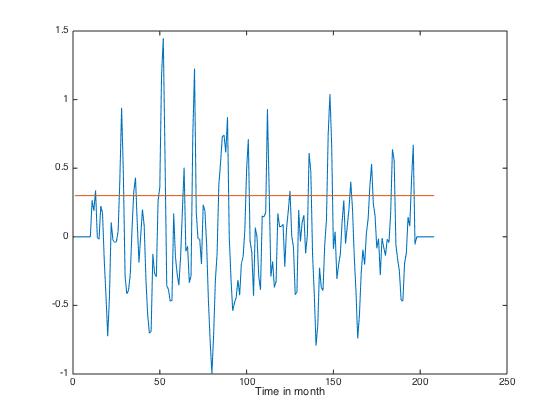}}}\\%
\subfigure[Time lapse between fluctuations]{
\resizebox*{9.1cm}{!}{\includegraphics{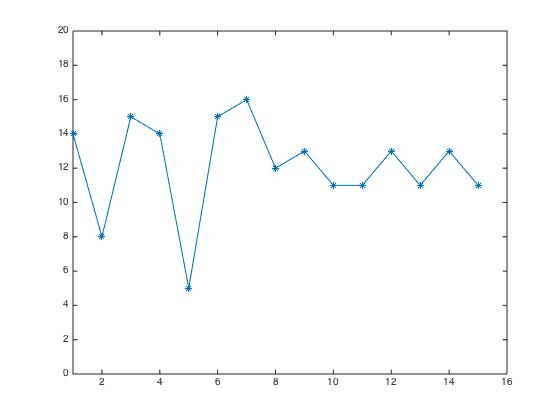}}}%
\caption{Tomatoes (1 kg)}%
\label{T}
\end{center}
\end{figure*}

\begin{figure*}
\begin{center}
\subfigure[Consumer prices in France (--blue) and its mean (--red)]{
\resizebox*{9.1cm}{!}{\includegraphics{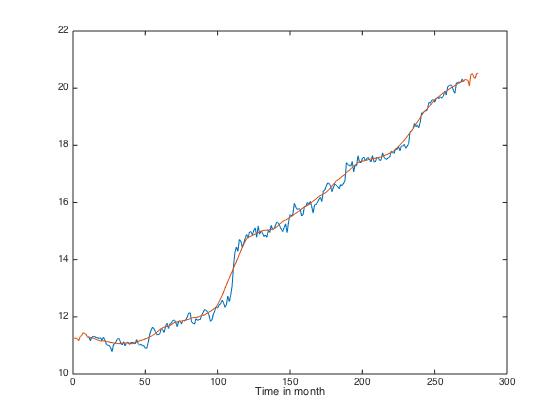}}}%
\subfigure[Fluctuations (--blue) and threshold (--red)]{
\resizebox*{9.1cm}{!}{\includegraphics{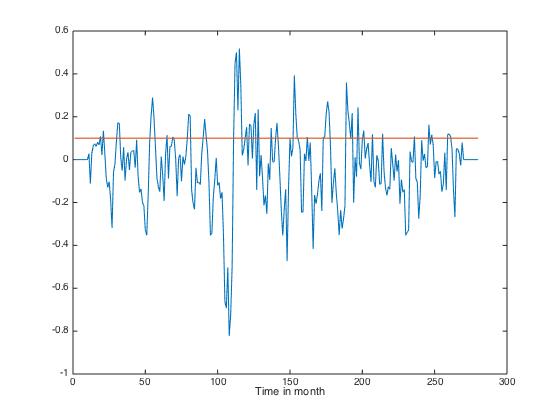}}}\\%
\subfigure[Time lapse between fluctuations]{
\resizebox*{9.1cm}{!}{\includegraphics{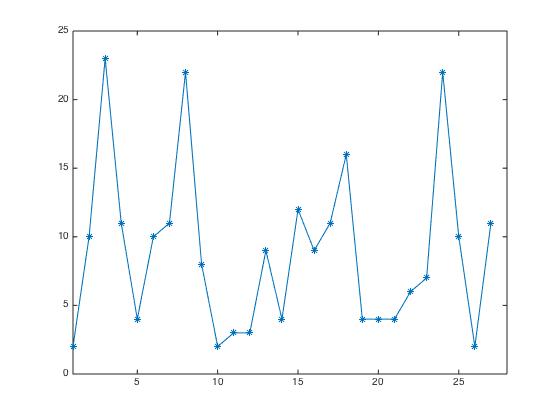}}}%
\caption{Lamb - middle neck (1 kg)}%
\label{V}
\end{center}
\end{figure*}

Select the following monthly data:
\begin{itemize}
\item Figure \ref{P}: apples from January 1998 to April 2015,
\item Figure \ref{T}: tomatoes from January 1998 to April 2015,
\item Figure \ref{V}: lambs from January 1992 to April 2015.
\end{itemize}
Equation \eqref{decomposition} is illustrated by the Figures \ref{P}-(a), \ref{T}-(a) and \ref{V}-(a). The means are computed via a centered moving average of length 20.
Figures \ref{P}-(b), \ref{T}-(b) and \ref{V}-(b) display the fluctuations. Strong seasonalities appear after some time in Figure \ref{T}-(c). There are only weak seasonalities in Figure \ref{V}-(c).

\subsection{Unemployement}
\begin{figure*}
\begin{center}
\subfigure[Job-seekers in France (--blue) and its mean (--red)]{
\resizebox*{9.1cm}{!}{\includegraphics{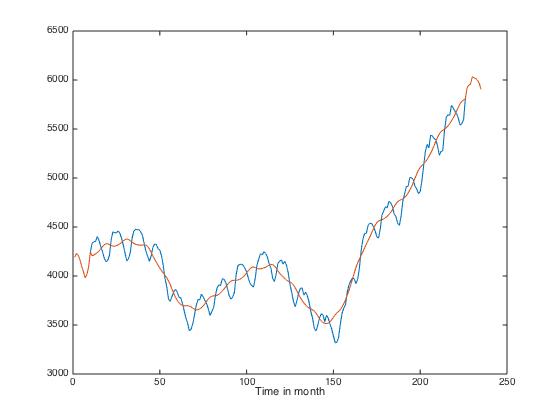}}}%
\subfigure[Fluctuations (--blue) and threshold (--red)]{
\resizebox*{9.1cm}{!}{\includegraphics{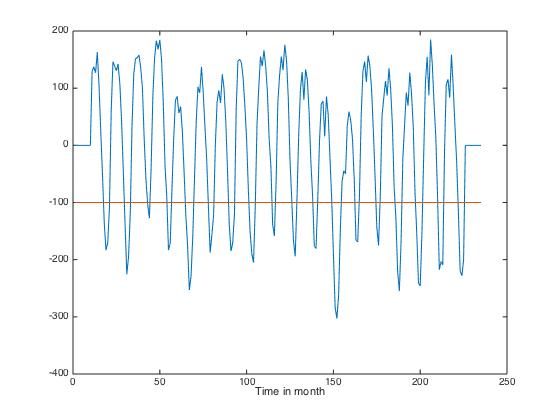}}}\\%
\subfigure[Time lapse between fluctuations]{
\resizebox*{9.1cm}{!}{\includegraphics{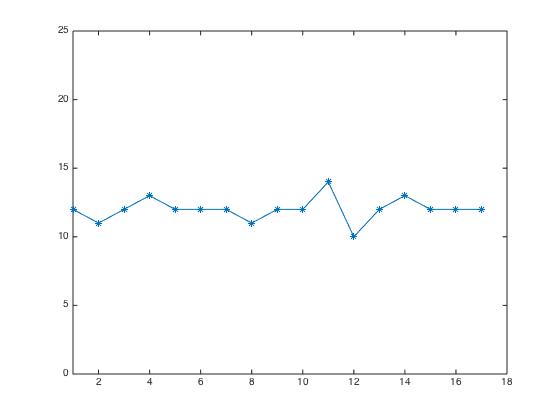}}}%
\caption{Unemployment}%
\label{C}
\end{center}
\end{figure*}

Figure \ref{C}-(a) displays the number of job-seekers from December 1995 to June 2015. The mean, which is estimated via a sliding window of length 20 months, and the threshold are shown in
Figure \ref{C}-(b)). A strong seasonality are obvious in Figure \ref{C}-(c).



\subsection{Air passengers}
\begin{figure*}
\begin{center}
\subfigure[Monthly number of passengers (--blue) and its mean (--red)]{
\resizebox*{9.1cm}{!}{\includegraphics{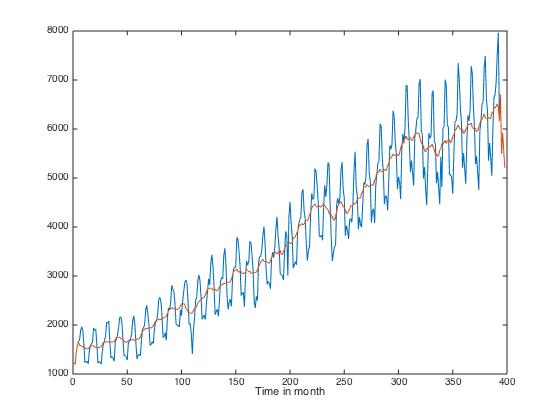}}}%
\subfigure[Fluctuations]{
\resizebox*{9.1cm}{!}{\includegraphics{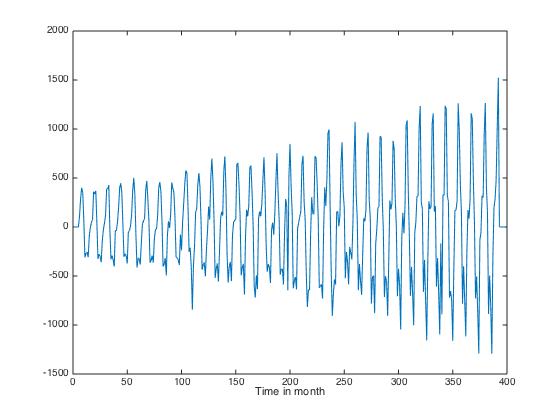}}}\\%
\subfigure[Normalized fluctuations (--blue) and threshold (--red)]{
\resizebox*{9.1cm}{!}{\includegraphics{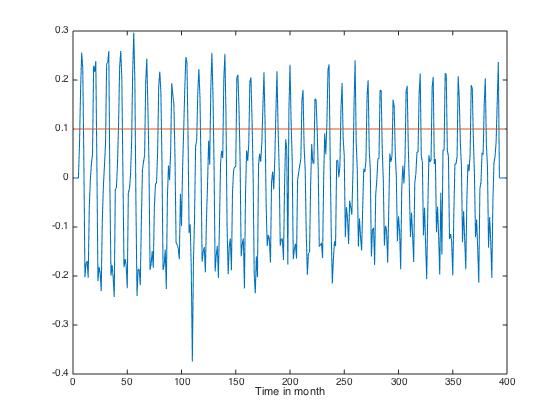}}}%
\subfigure[Time lapse between fluctuations]{
\resizebox*{9.1cm}{!}{\includegraphics{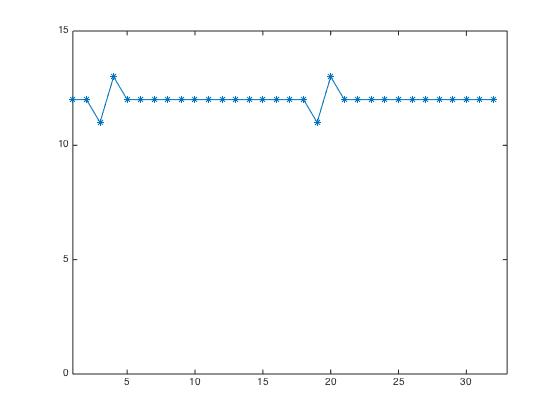}}}%
\caption{Air passengers}%
\label{Vol}
\end{center}
\end{figure*}

Monthly data from January 1982 to April 2015 are presented in Figure \ref{Vol}-(a). The mean is obtained via a centered 
moving average of length 10. Figure \ref{Vol}-(b) shows that the magnitude of fluctuations is related to the number of passengers. $\frac{X_{\tiny{\rm fluctuation}}(t)}{E(X)(t)}=\frac{X(t) -E(X)(t)}{E(X)(t)}$ is therefore presented in Figure \ref{Vol}-(c) as well as the threshold. Figure \ref{Vol}-(d) display a clear-cut strong seasonality.

\subsection{Internet traffic}\label{internet}
\begin{figure*}
\begin{center}
\resizebox*{9.1cm}{!}{\includegraphics{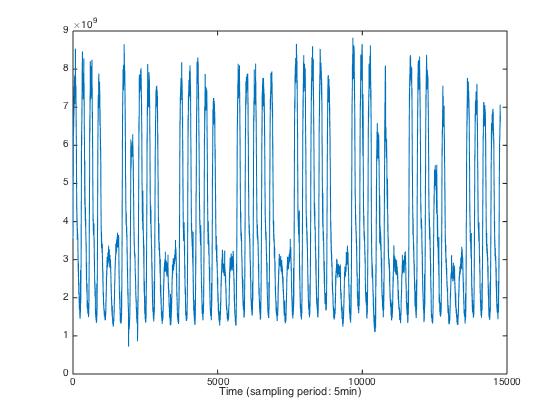}}%
\caption{Internet traffic per 5 minutes (original series)}%
\label{Io}
\end{center}
\end{figure*}

\begin{figure*}
\begin{center}
\subfigure[Internet traffic (--blue) and its trend (--red)]{
\resizebox*{9.1cm}{!}{\includegraphics{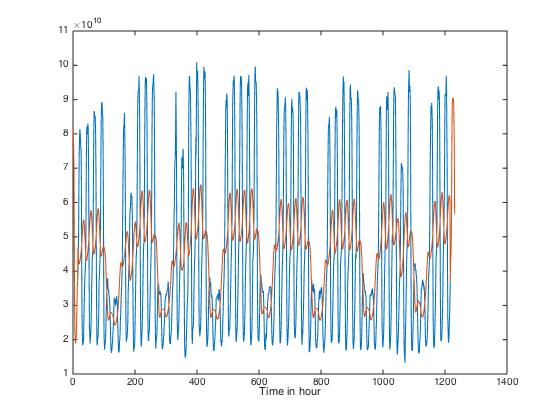}}}%
\subfigure[Fluctuations]{
\resizebox*{9.1cm}{!}{\includegraphics{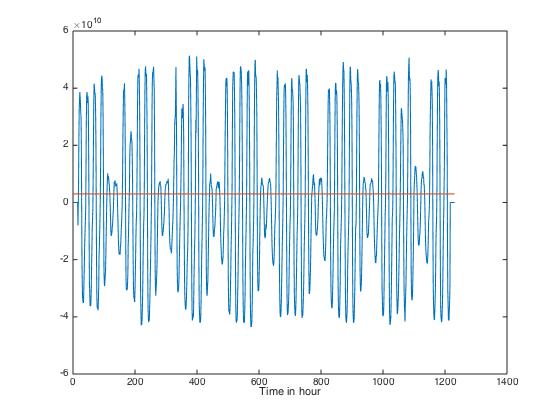}}}\\%
\subfigure[Time lapse between fluctuations]{
\resizebox*{9.1cm}{!}{\includegraphics{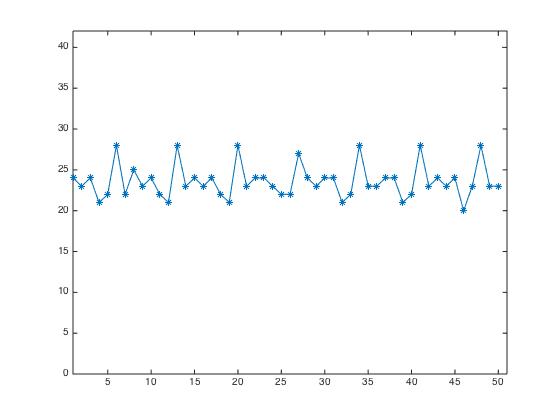}}}%
\caption{Internet traffic with a time scale of $1$ hour}%
\label{Ih}
\end{center}
\end{figure*}

\begin{figure*}
\begin{center}
\subfigure[Internet traffic (--blue) and its trend (--red)]{
\resizebox*{9.1cm}{!}{\includegraphics{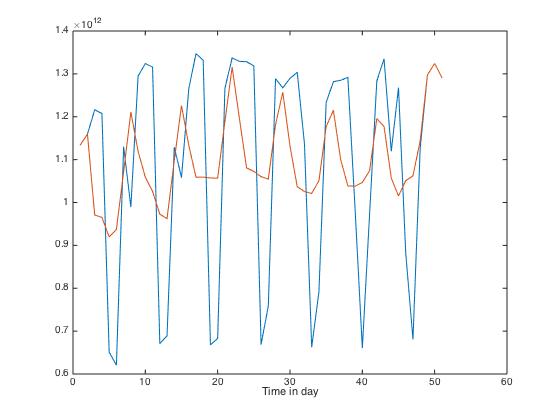}}}%
\subfigure[Fluctuations]{
\resizebox*{9.1cm}{!}{\includegraphics{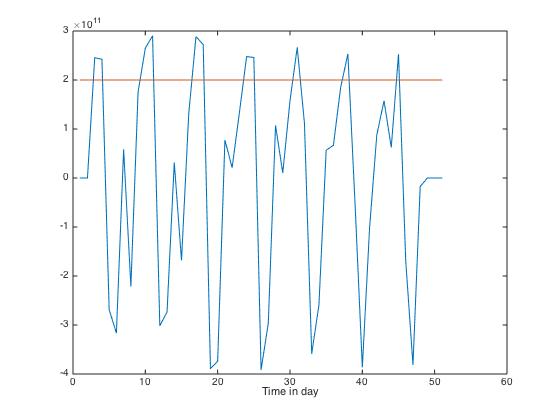}}}\\%
\subfigure[Time lapse between fluctuations]{
\resizebox*{9.1cm}{!}{\includegraphics{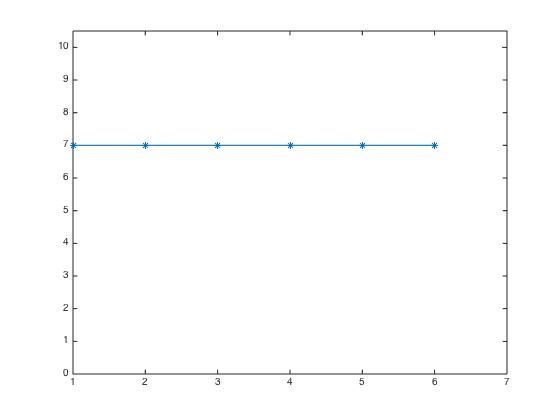}}}%
\caption{Internet traffic with a time scale of $1$ day}%
\label{Id}
\end{center}
\end{figure*}

Figure \ref{Io} shows internet traffic data, in bits, which were collected every $5$ minutes from 06:57 a.m., 7 June 2005,  to 11:17 a.m., 31 July 2005.
Several time scales are used via Equation \eqref{sumcumulee}.
Figures \ref{Ih}-(b) and \ref{Ih}-(c) display respectively the fluctuations and the seasonalities with respect to a time scale of $1$ hour. The trend in Figure \ref{Ih}-(a) is computed via a moving average of 30 points. Figure \ref{Id} presents the same type of results with a time scale of $1$ day. 

It should be clear that the above seasonalities have been detected and extracted only thanks to several time scales.

\section{Conclusion}\label{conclusion}
The above encouraging results need of course to 
\begin{itemize}
\item be confirmed by several other concrete case-studies, 
\item yield forecasting improvements.
\end{itemize}
It would be also most rewarding to see if our viewpoint may be helpful 
for investigating some complex questions like, for instance, the famous \emph{Kondriatev waves} (\cite{kondra}\footnote{Kondratjew is a German spelling of this Russian name.}), which are quite debatable (see, \textit{e.g.}, \cite{boss}). Are the time series which are at our disposal 
long enough to start such a research? A recent paper by \cite{koro}, where techniques stemming from spectral analysis are used, looks quite stimulating.


\bibliography{ifacconf}             

\end{document}